# Towards Cryogenic AI Hardware: Over 99% Precision In-Memory Computing at 4 K with 3-bit Ferroelectric $Hf_{0.5}Zr_{0.5}O_2$ Capacitors

*Ella Paasio[1], Simon Thomann[2], Rikhard Ranta[1], Ruhan Kartik[1], Sozaraj Rasappa[1], Andrea Padovani[3], Hussam Amrouch[2], Sayani Majumdar[1*]*

[1]Faculty of Information Technology and Communication Sciences, Tampere University, Finland

[2]School of Computation, Information and Technology, Technical University of Munich, Germany

[3] Department of Engineering Sciences and Methods (DISMI), University of Modena and Reggio Emilia, Italy

* Corresponding author, e-mail: sayani.majumdar@tuni.fi

***Abstract***

High-accuracy cryogenic hyperdimensional computing (HDC) is demonstrated using $Hf_{0.5}Zr_{0.5}O_2$ (HZO) ferroelectric capacitors (FeCAPs) as multilevel analog memories for capacitive compute in-memory (CIM) architectures. On the ISOLET spoken-word dataset, FeCAP-stored vectors achieve up to 92 % classification accuracy at 4 K, while charge-domain matrix– vector multiplication (MVM) exhibits near-ideal behaviour with linearity of 1 and error probability below 1% for 4×4 and 4×8 configurations, operating below 3 V. Unlike the performance at 300 K, HDC and CIM at 4 K remains resilient to electric-field-cycling-induced fatigue, thanks to enhanced cryogenic device characteristics, including reduced state variability, strongly suppressed leakage current and stable retention, as well as 78% faster switching. Furthermore, these multibit FeCAPs exhibit endurance beyond $10^9$ cycles with polarization degradation below 8%, preserving multilevel integrity. Benefits can also be observed at 300 K, following cooldown and wake-up pulsing at 4 K and subsequent warming up. These results establish multibit FeCAPs as a highly reliable, ultra-low-energy, and accurate analog CIM platform for future cryogenic AI-computing hardware.

## Introduction

Compute-in-memory (CIM) is a computing paradigm that performs matrix-vector multiplications (MVMs) and logic operations directly within memory arrays, minimizing data movement between memory and processors. By reducing data shuttling between processor and memory, CIM achieves substantial energy savings and high computational throughput. At cryogenic temperatures, where dissipated energy by the electronic components can become prohibitively expensive, [1], [2] reducing data communication overhead is even more critical than improving computational efficiency itself. [3], [4] Cryogenic compute in-memory (Cryo-CIM) extends the CIM paradigm to cryogenic environments, typically operating between 4 K and 77 K. By combining the inherent data-transfer efficiency of CIM with cryogenic electronics, Cryo-CIM is emerging as a key technology for overcoming hardware and data-transfer bottlenecks in large-scale quantum computing systems, advanced AI accelerators, and deep-space electronics. The applications and operating principle of Cryo-CIM is schematically shown in **Fig. 1**.

Cryogenic memory operation has been demonstrated [3], [5] from standard CMOS memories like static random-access memory (SRAM), dynamic random-access memory (DRAM) and Flash to emerging memories including phase change memories (PCMs) [6] and ferroelectrics (FEs) [7]. It is seen that cryogenic operation improves endurance and retention, however at the cost of higher voltage requirements for operation.

Cryogenic CIM performance is not so widely studied. The performance reported from a few selected systems are benchmarked in **Table I**. However, most reports show the feasibility of cryogenic operation with focus on metrology and noise [3], [9] rather than investigating computing accuracy. Fe-FinFET based ternary content addressable memory (TCAMs), used for hyperdimensional computing (HDC) at 10 K, showed degradation of computing accuracy at cryogenic temperatures, due to increased memory window (MW). [7] In this work, we report ferroelectric $Hf_{0.5}Zr_{0.5}O_2$ (HZO) ferroelectric capacitors (FeCAPs) where multilevel states are obtained both at 300 K and 4 K, however the reliability of intermediate states improve significantly at cryogenic temperatures. Also, the fatigue behaviour of the FeCAPs improved hugely that led to near-perfect CIM precision at 4 K using HZO-based FeCAP technology, with linearity of 1 and minimal error probability. Depending on the temperature of the wake-up and thermal cycling history, the total switchable polarization changes, however near unity linearity at 4 K remains identical that is a promising feature for high precision MVMs.

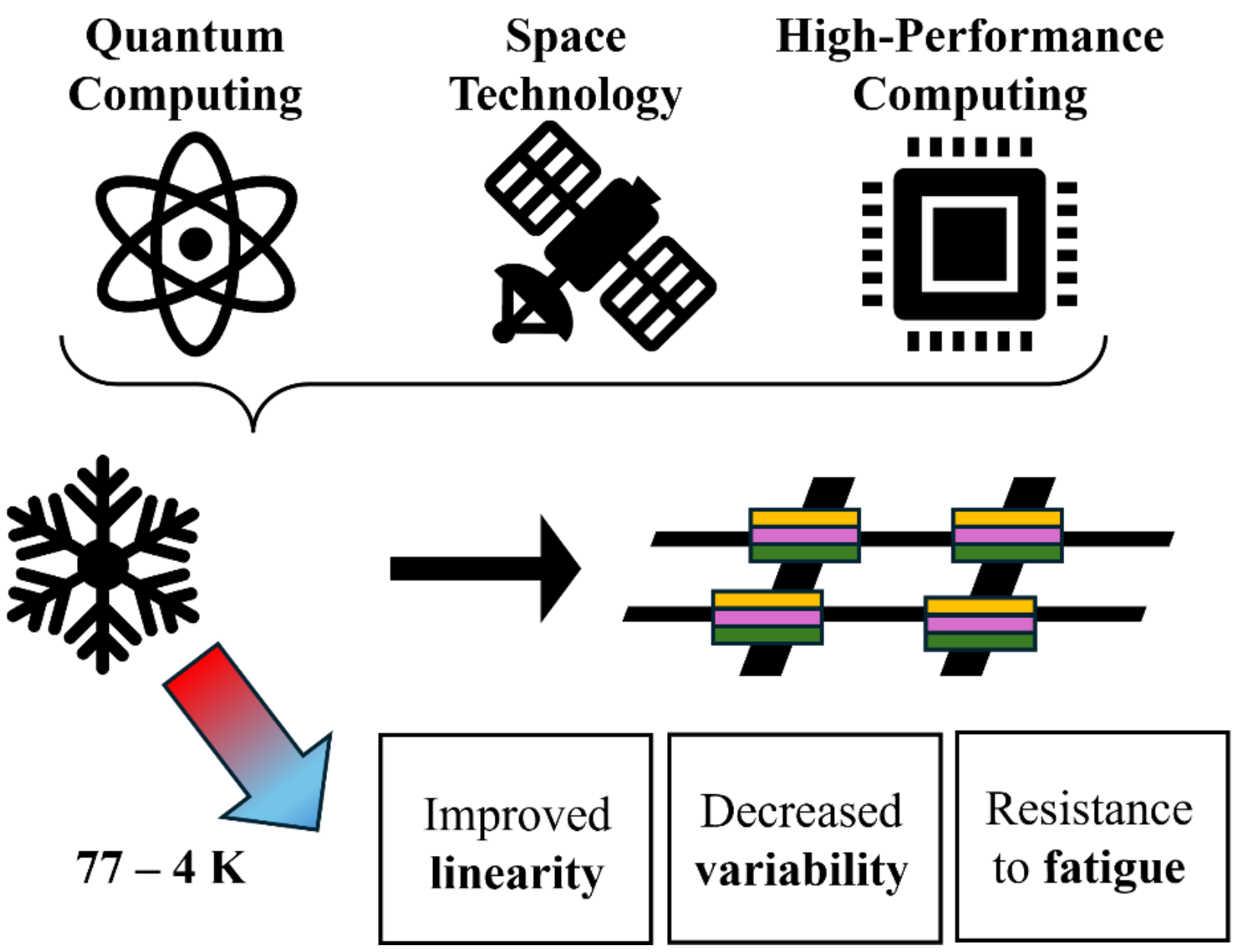

**Fig. 1**: Cryo-CIM aims at solving data-transfer bottlenecks, achieving substantial energy savings and high throughput, in cryogenic environments where dissipated energy is prohibitively expensive. Ferroelectric components show promising performance as a cryogenic memory element with more controlled domain dynamics that leads to improved linearity, decreased variability and better resistance to fatigue over prolonged bias stressing.

**TABLE I:** Benchmarking Cryo-CIM results from different technologies. Regime of cryogenic operation is 4-77 K.

| Memory | Op. in Cryo. | CIM | Endurance | Retention | Important KPIs | Ref. |
|---|---|---|---|---|---|---|
| SRAM, Cool-CIM | Excellent | Digital | Unlimited | Volatile | EDP: -54% energy efficiency at 77K: 308.5 TOPS/W | [3], [13] |
| eDRAM | Good | Digital | High | Volatile | Energy efficiency at 4K: 603 TOPS/W | [14] |
| SRAM, Cryo-RNN | Good | Digital | High | 9.1 s at 4K | EDP: 52.2 µs.nJ, QEC at 4K | [15], [16] |
| PCM | Fair | Analog | $10^8$ | Nonvolatile with drift | MVM error at 5K: >10%, stable analog computation at 5 K | [6] |
| Fe-FinFET | Fair | Digital | $10^5$-$10^6$ | Nonvolatile | Decreased accuracy for HDC at 10 K compared to 300 K | [7] |
| FeCap | Excellent | Analog | $>10^9$ | Nonvolatile, $> 10^5$ s | MVM error at 4 K 0.7%, Linearity: 1 (2×3-bit), 1.1 TOPS/W. HDC at 4 K with 92 % accuracy (20 states) | This work |

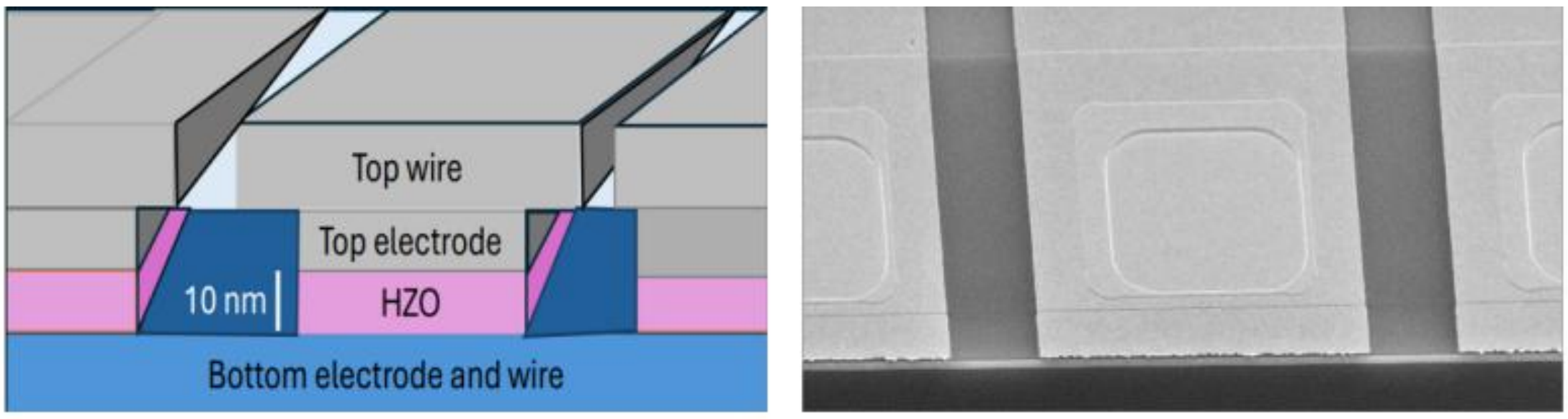

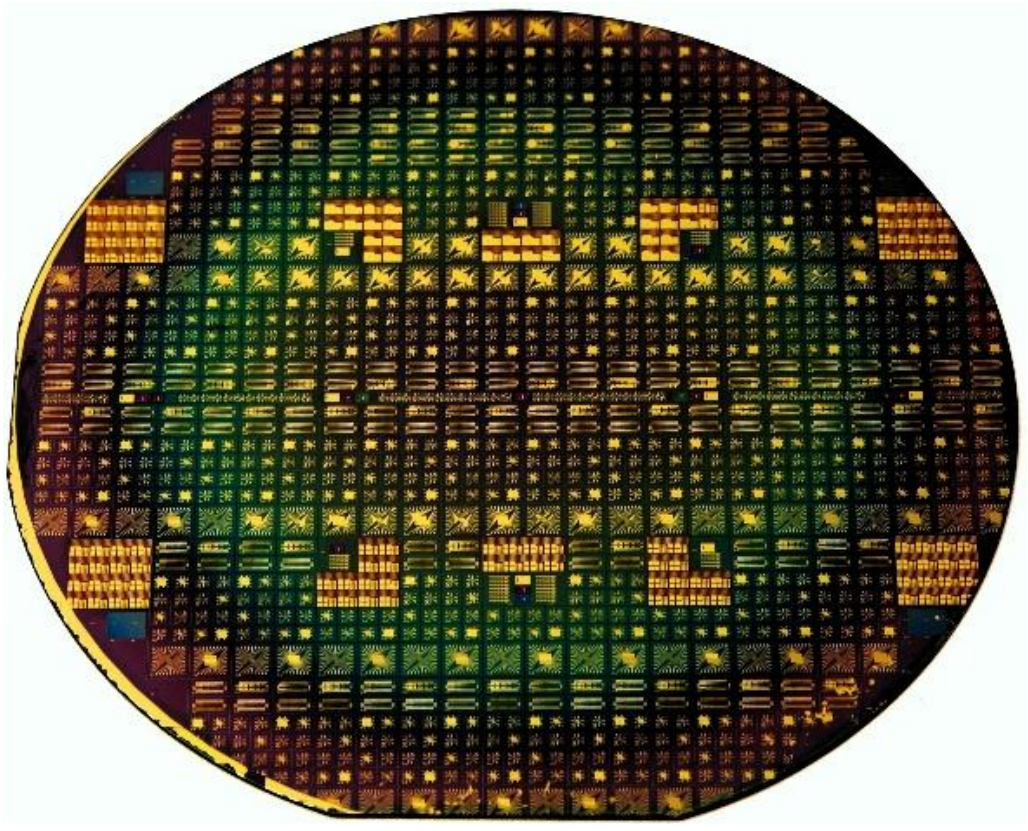

- $SiO_2$ deposition on Si substrate
- PEALD deposition of TiN (30 nm)
- Thermal ALD of HZO (10 nm) and $Al_2O_3$ (1 nm)
- Patterning for Top Electrode definition
- Ti/Au evaporation and Lift-off
- Patterning for bottom wire definition
- Etching of HZO and TiN layers
- PECVD of deposition of passivation layer of $SiO_2$
- Via opening
- Top wire evaporation of Ti/Au
- Rapid thermal annealing at 400 $^{o}$C in $N_2$ atmosphere

**Fig. 2:** (Top panel) Schematic diagram of the device stack and scanning electron microscopy (SEM) image the fabricated FeCAP devices together with the whole wafer image and the fabrication process flow.

## Experimental

FeCAPs were fabricated following the process illustrated in **Fig. 2** and characterized with Aixacct 2000E and Keysight B1500A after placing the samples inside the Advanced Research System ARS-PS-CC10 cryogenic probe station. Details of device stack and it's structural and electrical properties are presented in [8]. In these experiments, devices were woken up with $10^3$ bipolar 1 kHz square pulses at 3 V at mentioned temperatures. FE properties were characterized using triangular Positive-Up Negative-Down (PUND) measurements at 1 kHz. DC static current – voltage measurements and capacitance-voltage measurements were performed with Keysight B1500 semiconductor parameter analyser. Endurance measurements were performed after pulsing the devices with a certain voltage amplitude at 100 kHz frequency, taking P-V loop measurements after each decade of pulsing. Retention was measured by applying a triangular full PUND pulsing to obtain the initial polarization. After that, two PUND-like pulses were applied unipolarly to obtain how much ferroelectric switching can happen, which is ideally zero due in case of full retention. The obtained switching minus non-switching pulses were then subtracted from the initial measurement to represent the drop of polarization during retention. These unipolar reading pulses were applied at increasing time between pulses, and the time to next pulse was used. Temperature dependence was measured after cooling down the samples to 4 K initially followed by warming up to 300 K, taking data at intermediate temperature ranges. Distributions of multi-bit states were extracted with data from 10 pulses for 20 intermediate states from 4 K to 300 K. CIM experiment details are provided in the Compute-in-Memory section.

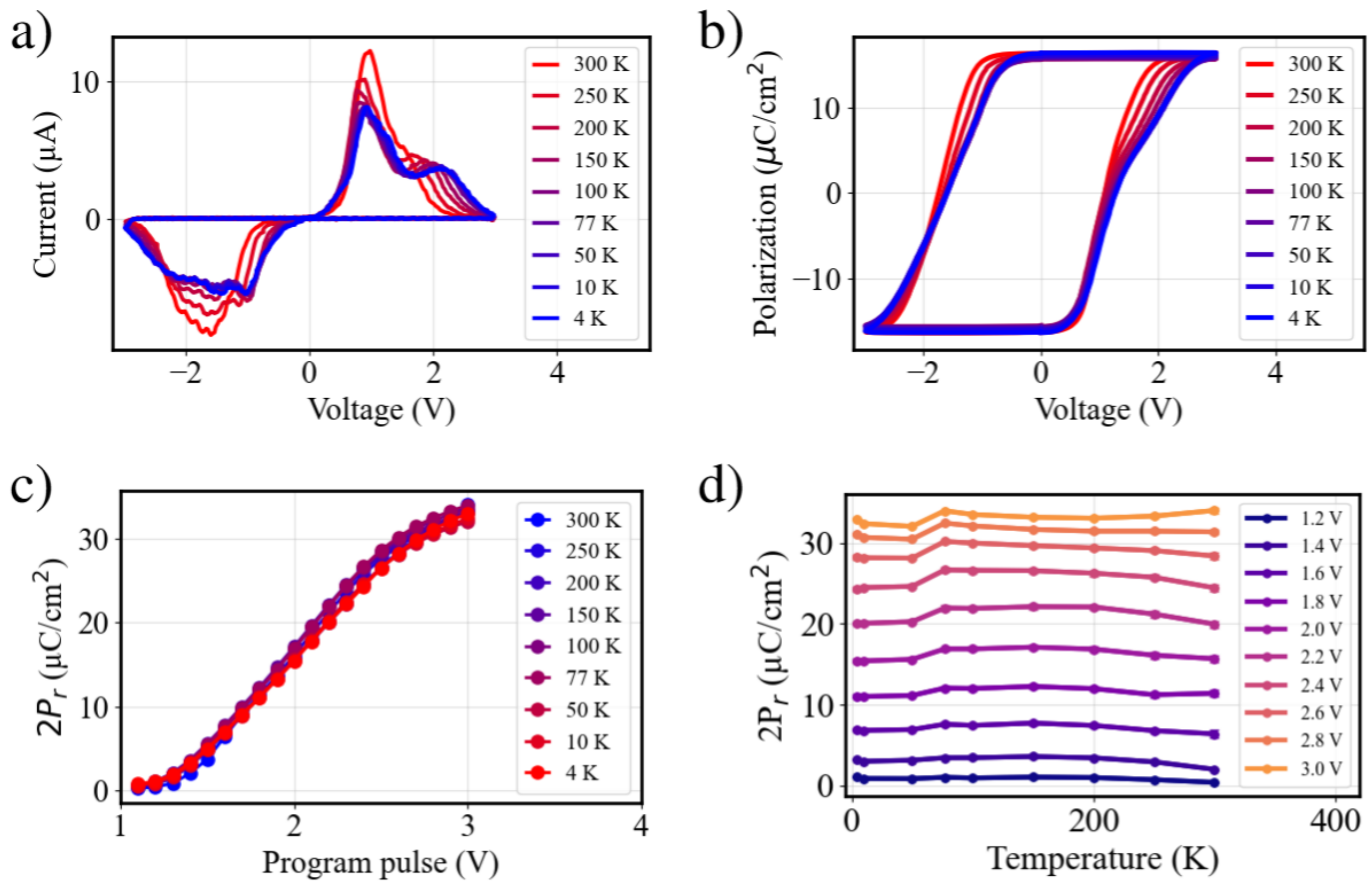

**Fig. 3**: a) $I - V$ and b) resulting $P - V$ loops show remarkable stability across cryogenic temperatures to room temperature with program voltage of 3 V, when woken up in 4 K. c) Remanent polarization ($P_r$) against different pulse amplitudes reveals good analog state linearity between 1.2 V and 3 V. d) The analog states are retained throughout temperature change. Largest difference observed around 77 K for the states with higher program voltage.

## Results and Discussion

### A. FeCAPs operation down to 4 K

The dynamic current-voltage ($I - V$) and polarization-voltage ($P{-}V$) curves of the HZO FeCAPs are shown in **Fig. 3 a)** and **b)**, respectively where remarkable temperature stability is observed between 4 K – 300 K when wake-up pulsing is applied at 4 K. Only with decreasing temperature, the shape of the $I$ -$V$ loop changes from a sharp ferroelectric polarization switching related peak to a more gradual and multiple peak-like transition. The analog programmability of polarization states is stable over the full range of voltages and temperatures, as shown in **Fig. 3 c)** and **d)**. In our previous work, [17] through Jiles-Atherton modelling of experimental FeCAP cryogenic performance data and predictive modelling of the gradual polarization switching, we found that linearity of intermediate states can be better at low temperatures due to less abrupt domain reversal at 4 K. Here, we confirm this prediction experimentally.

The intermediate state standard deviation ($\sigma$) decreases at low temperature, and 8 states remain within $2\sigma$ of the 300 K value even at 4 K (**Fig. 4 a)).** The switching time properties were analyzed after Cryo-WU, and we observed that the lower temperature allows for a faster ferroelectric switching. The improvement in switching time is dependent on the program voltage amplitude, and with 3 V the switching time is 78.2 % lower in 4 K compared to 300 K, which is shown in **Fig. 4 d)**. This is promising for high bandwidth memory operation in 4 K. A more detailed description of these results can be found in Ref. [9]. As expected, thermally activated leakage current through HZO becomes significantly diminished at low temperatures (**Fig. 4 c**)), which corresponds towards lower static energy consumption and better reliability in general. At 2.5 V, the current density is 4.13 pA/$\mu m^2$ in 300 K, which drops down to 0.059 pA/$\mu m^2$ in 4 K, which is in the lower end of the resolution of the measurement unit. Leakage current drop of 98 % was measured. In previous reports of HZO-based CIM appointed worse cryogenic operation to increased voltage requirements [7]. Here, we observed that when woken up in 4 K, there is minimal change to the coercive field and memory window values when comparing the sample in 4 K and in 300 K (**Fig. 4 c**)). The difference in MW, defined as $\Delta\mathrm{MW}(\mathrm{V_{prog}}) = \frac{MW_{4K}(\mathrm{V_{prog}}) - MW_{300K}(\mathrm{V_{prog}})}{MW_{300K}(\mathrm{V_{prog}})}$ observed is dependent on the program voltage, and the largest difference observed is at very low voltages of 1 V, where ΔMW is within 15 % between 4 K and 300 K.

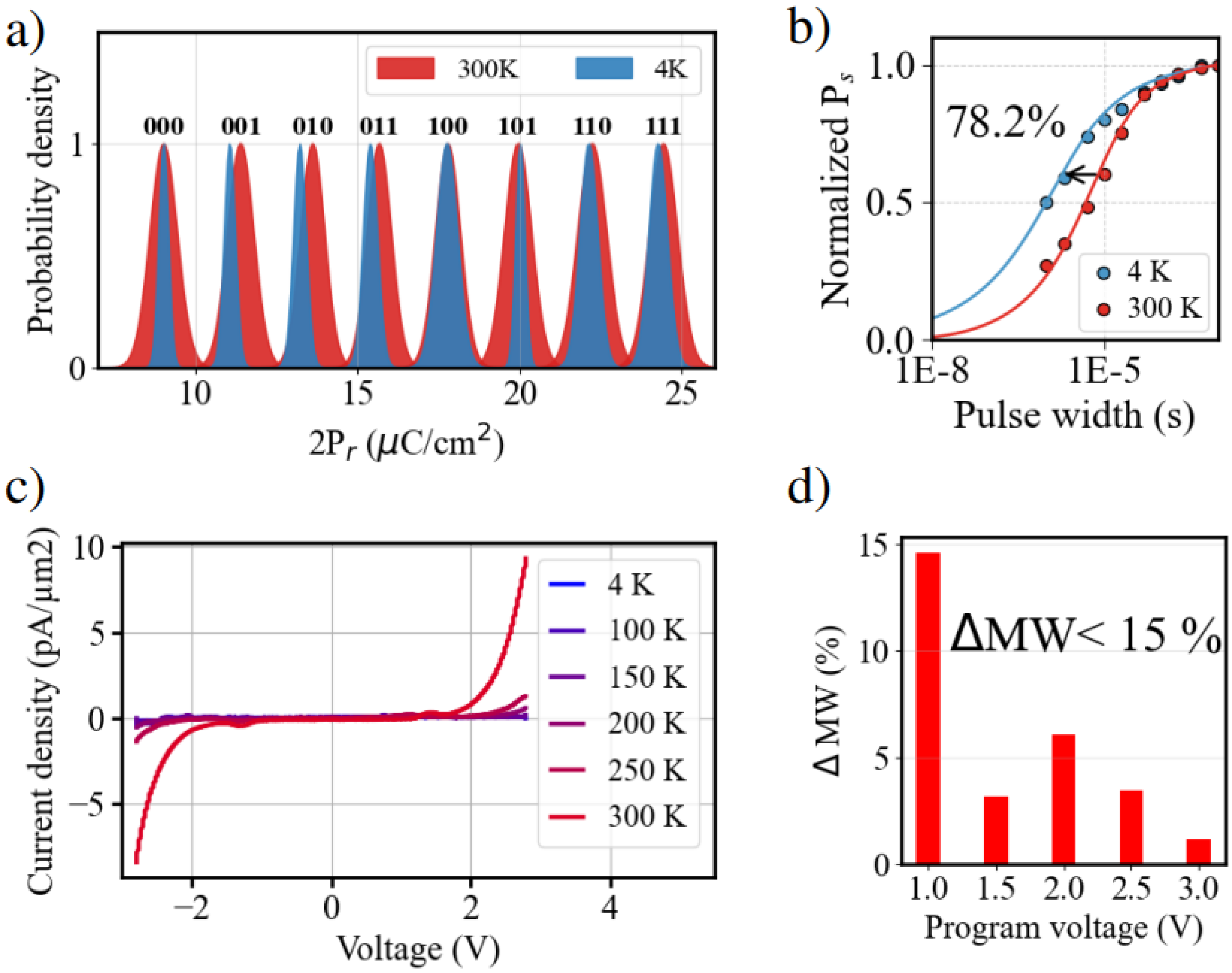

**Fig. 4:** a) Memory state distributions as Gaussian fits show analog state variability. b) Improved switching speed characteristics at 4 K. c) Cool-down to 4 K reduces leakage current significantly. d) MW is stable across temperatures. At 3 V programming, MW is 2.8 V. At 1 V operation, MW=1.4 V. Change in MW due to temperature from 4 K to 300 K.

Major impact on analog programmability is observed depending on wake-up (WU) temperature. WU at 300 K yields 28% higher $P_r$ compared to that at 4 K and then heated to 300 K (**Fig. 5 a**)). WU at 300 K followed by cool-down to 4 K exhibits high linearity, but with increased voltage requirement (**Fig. 5 b**)). This effect is not well understood yet, but given the nature and physics of WU, it is hypothesized to be related to oxygen vacancy redistribution, domain pinning or phase transformation under combined effect of bias and temperature. Normalized analog linearity, the main contributor for CIM precision, is improved when WU pulsing was applied at 4 K. This trade-off between magnitude of polarization charge and weight update linearity highlights the importance of WU temperature optimization for accurate Cryo-CIM performance.

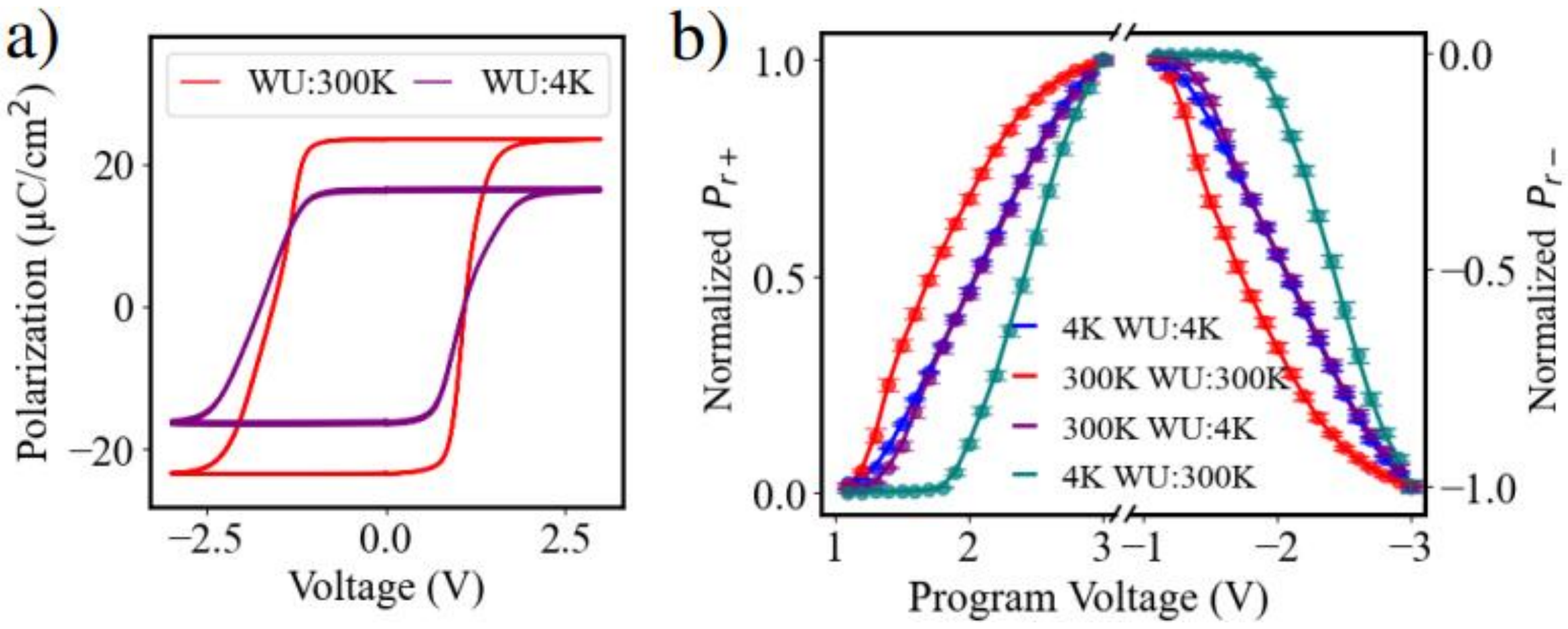

**Fig. 5**: a) P −V loops measured at 300 K after WU at 300 K and 4 K. Room temperature WU yields 28 % higher Pr. b) Normalized $P_r$ with respect to program voltage show intermediate state linearity is majorly impacted by both temperature and WU temperature.

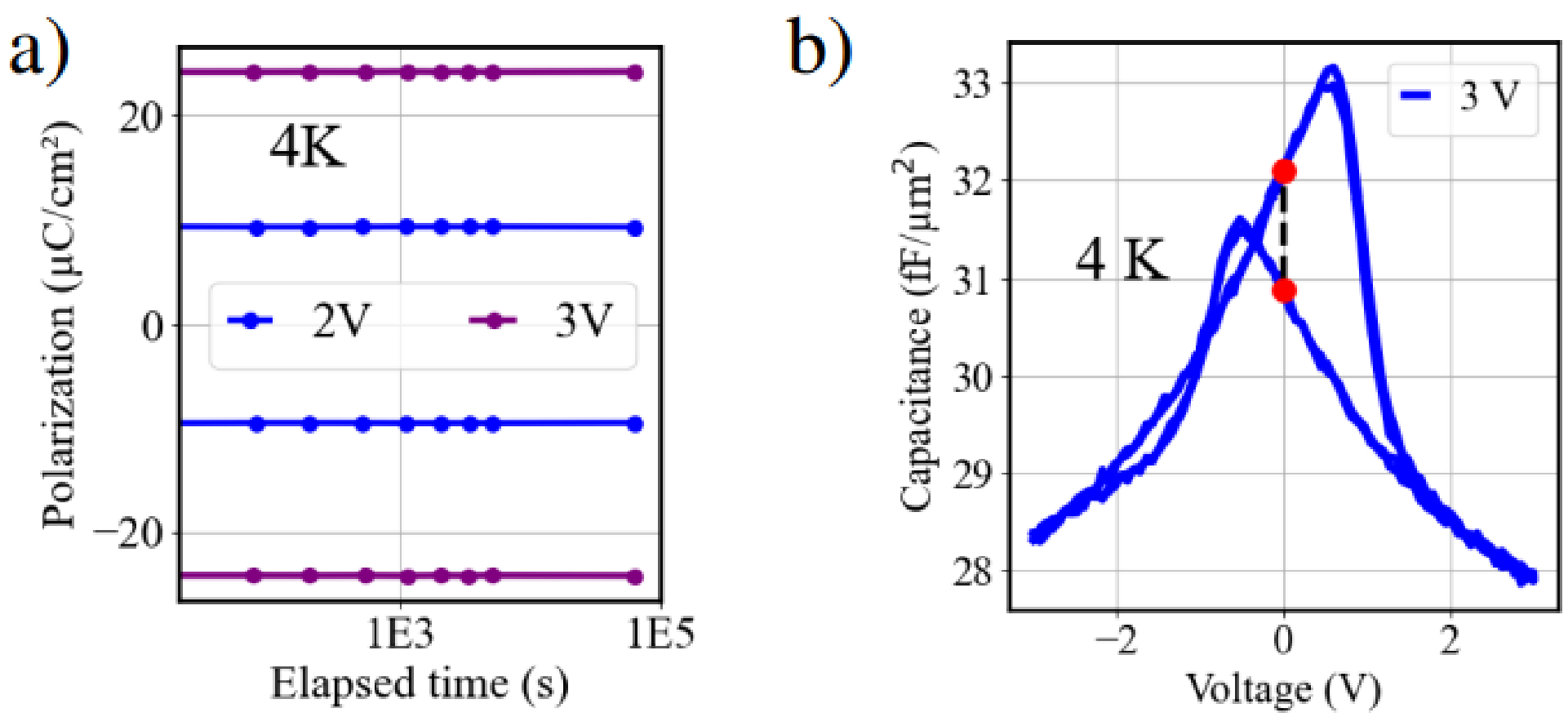

**Fig. 6**: a) Polarization retention shows no degradation at 4 K over the measured $10^5$ seconds for two different polarization states. b) Open capacitive memory window at 4 K at zero bias when scanned between ± 3V at 1 kHz frequency.

Polarization retention measurements performed for both 2 V and 3 V write operations show different polarization states can be retained for over $10^5$ s without any degradation as seen in **Fig. 6 a)**. The capacitance – voltage (*C-V*) measurements performed in 4 K after WU in 4 K showed an asymmetric butterfly-loop exists that leads to an open capacitive MW in 4 K as seen in **Fig. 6 b)** with 1 kHz measurement**.**

Endurance at 4 K is beyond $10^9$ (**Fig. 7a**)) when cycled with 3 V pulses with 100 kHz frequency, and $\Delta P_r$ < 8% while at 300 K devices experience hard breakdown near $10^7$ pulsing cycles. Intermediate states endurance tested at 4 K also show absence of fatigue (**Fig. 7 b**). Linearity of the intermediate states degrades significantly at 300 K after $10^7$ pulsing cycles (**Fig. 7 c**) while the linearity remains constant over $10^7$ pulsing cycles of the programming voltage at 4 K (**Fig. 7 d**)), which indicates mitigation of FE fatigue under cryogenic conditions. This is essential for cryo-FRAM technology where frequent write operations are essential.

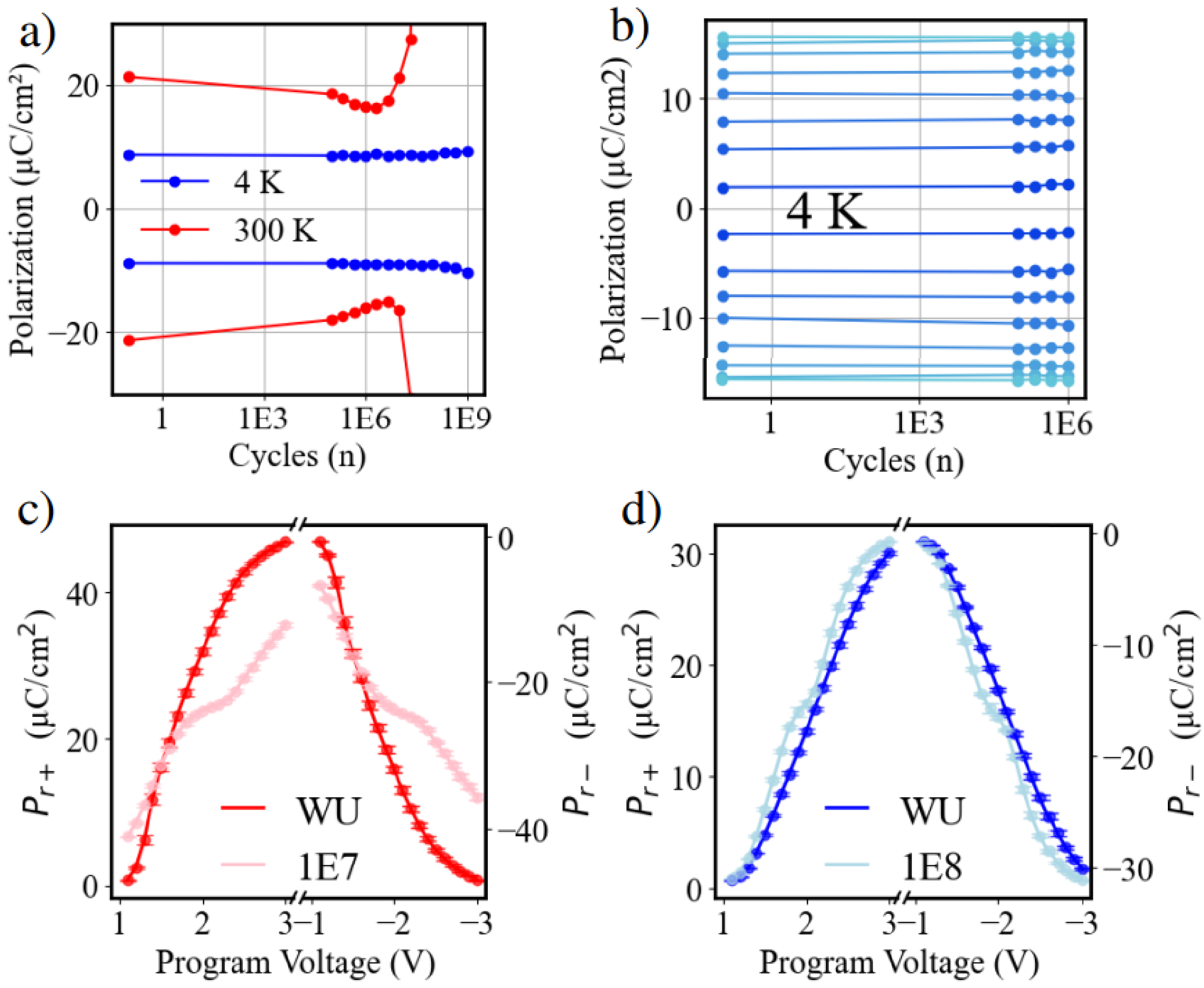

**Fig. 7**: a) Endurance cycling with 2 V at 300 K and 4 K, > $10^9$ pulses at 4 K. b) Endurance of intermediate states from 1 V to 3 V. c) Analog state linearity and polarization at 300 K degrade with $10^7$ electric field cycles of 2 V amplitude. d) No significant degradation due to $10^8$ pulses of 2 V amplitude at 4 K, but linearity is influenced.

Improvement of both retention and endurance properties at cryogenic temperature has been reported before in literature. [18, 19] In $HfO_2$-based ferroelectrics, it is known that at room temperature, electric field stress causes oxygen vacancies to migrate and accumulate at grain boundaries or interfaces, leading to domain pinning, leakage currents, and breakdown. [20] As shown in **Fig. 4(c)**, cooling down leads to freeze-out of thermal activation energy making oxygen vacancy redistribution much more difficult, resulting in significant decrease in leakage currents.

This decreased leakage results in much more stable operation of FeCAPs at low temperature. In the current experiments, we see that

**B. Compute in-memory (CIM)**

The analog state properties are evaluated as capacitive memory elements for charge-based CIM, where polarization modulated capacitance ($C$) and read voltage ($V$) generate charge outputs ($Q = V \times C$). Analog multiplication precision is evaluated within the most linear operating range using Monte Carlo-sampled $V - C$ products, digitized by an analog-to-Digital converter (ADC) with equidistant decision boundaries. The resulting confusion matrices (**Fig. 8 a**)) show improved 8×8 product precision and better fatigue resistance at 4 K compared to 300 K. **Fig. 8 (b)** shows analog multiplication precision metrics (def. in [10]) with different state configurations, and the impact of temperature and fatigue on those. At 4 K, ideal multiplication behaviour is obtained with 4 × 4 and 4 × 8, with linearity of 1 and error probability of <1%. Operand bit-width increases error probability. Extensive electric field cycling decreases the multiplication precision, however less at 4 K than at 300 K. For example, 4×4 error probability increases at 4 K by 8.5 %, when at 300 K it increases by 19.3 %. The 4 K implementation is consistently more precise even after fatigue than the 300 K implementation right after wake-up.

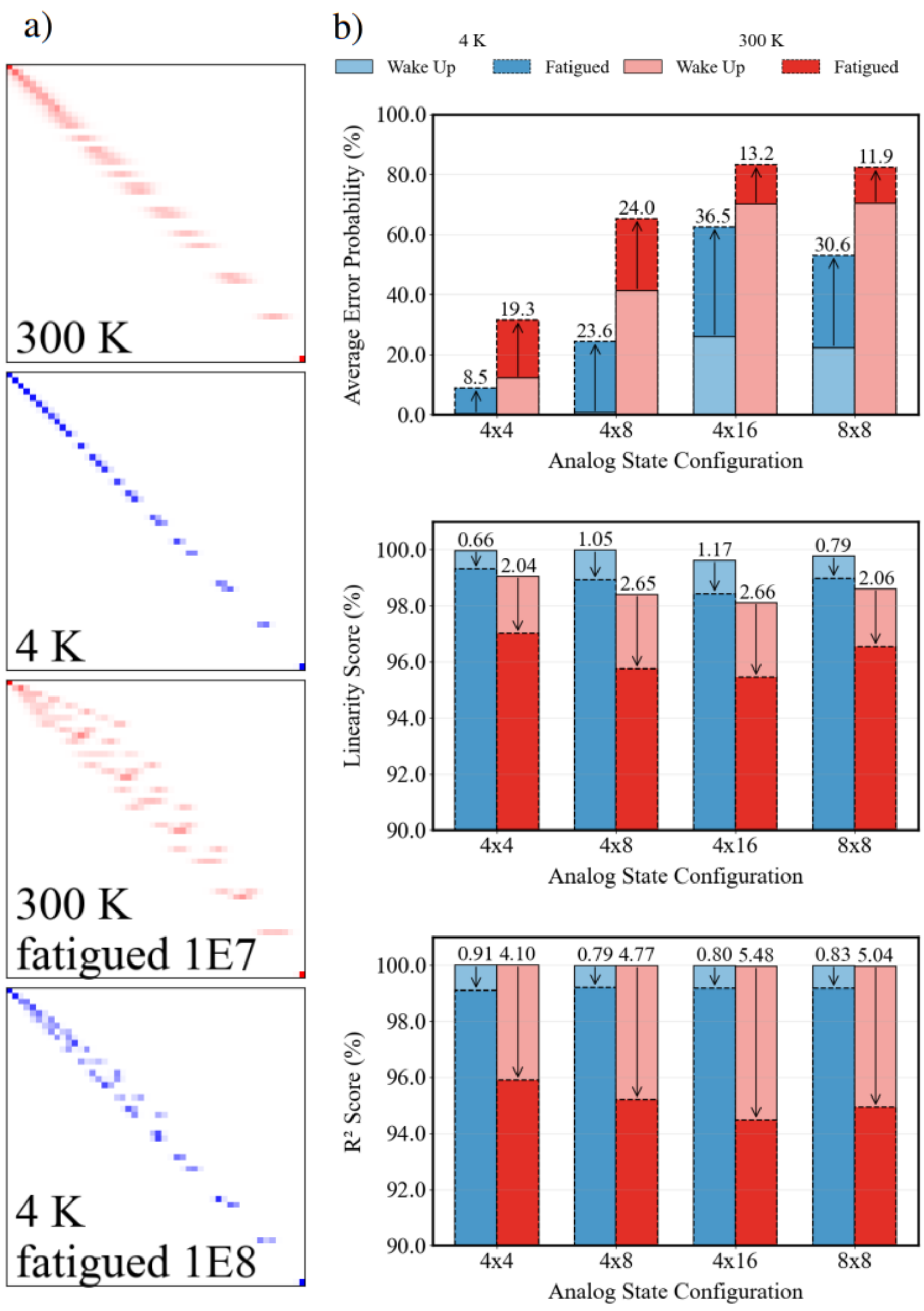

**Fig. 8**: a) MAC results of 8 × 8 (V state number × FeCAP state number), where the mathematically correct result is on the diagonal. b) Precision metrics of average error probability, linearity score and $R^2$ -score at 4 K and 300 K. The precision degradation due to fatigue in percentage points. Wake-up was done at operation temperature. All metrics are consistently better at 4 K, even after fatigue, than at 300 K after WU.

**Fig. 9 (a)** shows variability ($\sigma/\mu$ of $2P_r$) increases as the device heats up. This change in variability influences the analog multiplication accuracy, which is plotted for 4×8- multiplication against temperature in **Fig. 9 (b),** where cooling down improves each accuracy metric. We observe that WU at 4 K compared to room temperature decreases the error probability and improves linearity, but the $R^2$ -score suffers. This highlights the large impact WU temperature can have on the device properties, which carry out to analog multiplication.

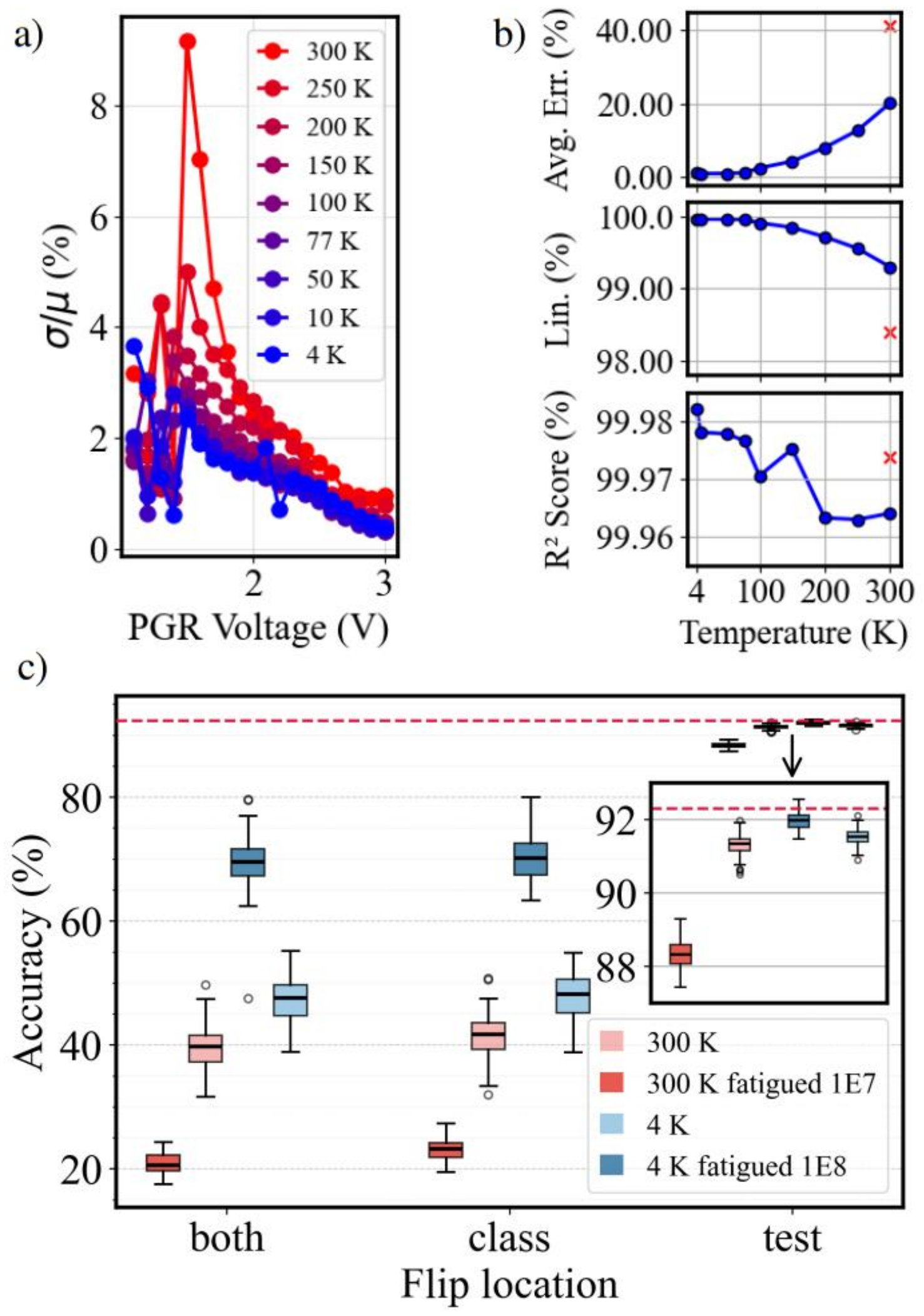

**Fig. 9:** (a) Following 4 K WU, σ/μ of $2P_r$ increases with temperature. (b) Precision metrics of analog multiplication against temperature, following 4 K WU in blue. At 4 K, all metrics are near-ideal. Red markers show precision metrics with WU done at 300 K. (c) HDC classification accuracy on the ISOLET dataset. Programming errors are injected as bit flips into class and/or test vectors. As test vectors, classification accuracy is 91 % both at 300 K and 4 K. At 300 K, accuracy is degraded by fatigue, whereas at 4 K, the accuracy improves due to tightened analog state distributions.

At 4 K, capacitance of the devices with area A = 50 × 50 μm$^2$ is ~3 pF. With a readout voltage of 0.5 V, reading a memory state consumes $E = \frac{1}{2}CV^2 \approx 0.37$ pJ/device. The program energy is $A \int V\, dP = 1$ pJ/cell at maximum program voltage. With 8-bit SAR ADCs, crossbar size 32 × 32, the power consumption is dominated by the ADCs, resulting in 1.1 TOPS/W for 2 × 2 multiplication. Further development in cryogenic ADCs is needed for full hardware-level CIM demonstration.

We evaluated FeCAPs for HDC classification, where information is encoded in high-dimensional distributed representations, making the approach highly resilient to noise, device variability, and memory errors. Programming errors were injected as bit flips into class and/or sample hypervectors of the ISOLET dataset [11] using fixed-point HDC [12] under the full programming voltage range (20 states), as shown in **Fig. 9 c**). FeCAP-stored test vectors achieved 91 % inference accuracy at both 300 K and 4 K. Fatigue reduced accuracy at 300 K but improved it at 4 K by decreasing the variability (σ) of higher-voltage states, enhancing error resilience. At full programming voltage, class-vector inaccuracies became dominant due to a large state space. When both test and class vectors were stored in FeCAPs, classification accuracy is decreased, however after continuous electric field cycling, the accuracy at 300 K falls to 20.8 % while at 4 K is improved to 69.4 %.

## Conclusions

In conclusion, in the current work we report operation of HZO-based FeCAPs down to deep cryogenic temperature of 4 K and evaluate their performance in analog compute-in-memory circuits. Compared to previous results, our results show that HZO FeCAPs at cryogenic temperatures can enable highly reliable and ultralow-energy analog charge-based CIM. At 4 K, improved linearity, reduced cycle-to-cycle variability between intermediate polarization states, suppressed leakage current, and mitigated fatigue upon $10^8$ pulsing cycles yield near-ideal matrix–vector multiplication performance and up to 92 % inference accuracy in hyperdimensional computing tasks. Polarization charge window of FeCAPs depends significantly based on the temperature at which the wake-up protocol is performed. Together with endurance beyond $10^9$ cycles, these results position HZO FeCAPs as a strong candidate for future cryogenic computing architectures.

## Acknowledgements

The work gratefully acknowledges financial support from KDT-ARTIC project (No. #101139908). ARCTIC project is supported by the Chips Joint Undertaking funded by European Commission, its members (including top-up funding by Belgium, Austria, Germany, Estonia, Finland, France, Ireland, The Netherlands and Sweden), the Canadian and the Swiss federal governments.